\documentstyle[epsf]{prhep97}

\makeatletter
\let\chapter\hid@chapter
\makeatother
\begin{document}

\thispagestyle{empty}
\flushright {\Large IEKP-KA/98-4}
\flushright {\Large February 20, 1998}

\flushleft

\vspace*{3cm}
\begin{center}
\huge{\bf
B  PHYSICS \\[0.5cm]
}
\Large Invited talk given at the International \\ Europhysics 
Conference on High Energy Physics, \\ 
19-26 August 1997, Jerusalem, Israel \\[0.2cm]

\Large {\bf Michael Feindt} \\[0.2cm]
\large 
{Institut f\"ur Experimentelle Kernphysik \\ Universit\"at Karlsruhe \\
http://wwwinfo.cern.ch/$\tilde{\mbox{ }}$feindt  \\[5cm] }
\normalsize

{\large \bf Abstract} \\
{\large A review over recent experimental progress in the physics of the 
fifth quark is given.}

\end{center}


\authorrunning{M.\,Feindt}
\titlerunning{{\talknumber}: B physics}
 

\def\talknumber{7} 

\title{{\talknumber}: B Physics }
\author{Michael \,Feindt
(feindt@cern.ch.)}
\institute{ Institut f\"ur Experimentelle Kernphysik, Universit\"at Karlsruhe, Germany}

\maketitle

\begin{abstract}
A review over recent experimental progress in the physics of the 
fifth quark is given.
\end{abstract}
\section{Introduction}
The value of the mass of the fifth, the beauty-quark, around 5 GeV leads
to a special role of the b-hadrons. 
The most heavy quark, the top quark, is too heavy to build hadrons. This
is because it can decay by ``weak'' interaction into a real W-boson and
a b-quark. This decay occurs much faster than the
typical time needed to bind with an antiquark into a meson by the strong 
interaction. Thus, hadrons containing a b-quark are the heaviest
hadrons. On the other hand, the b-quark mass is much larger
than the typical scale of the strong interaction, $\Lambda_{QCD}$, responsible
for the binding of quarks into hadrons. This is the reason for the success
of Heavy Quark Effective Theory. 
A B-meson consisting of a heavy b-quark and a light antiquark thus resembles
lots of properties of the hydrogen atom. 

This review gives a short (due to space limitations) summary of the state
of the field. More extensive recent summaries of LEP results are found 
in \cite{roudeau} and \cite{schneider}, on B-decays in \cite{drell}.
Note also the companion theoretical talk at this conference \cite{neubert}. 
The transparencies of this
talk can be found in \cite{feindt}.

At this conference the first experimental verification of the running of the
b-quark mass has been presented \cite{DELPHIrunningbmass,talk514}. The 
relative 3-jet rate in $b\bar{b}$ events 
is compared to the relative 3-jet rate
in light quark events. A three jet topology originates from
gluon radiation off a quark line. This radiation is suppressed for
\setlength{\unitlength}{1mm}  
\begin{figure}[ht]
\begin{picture}(55,60)(0,0 )
\mbox{\epsfxsize6.0cm\epsffile{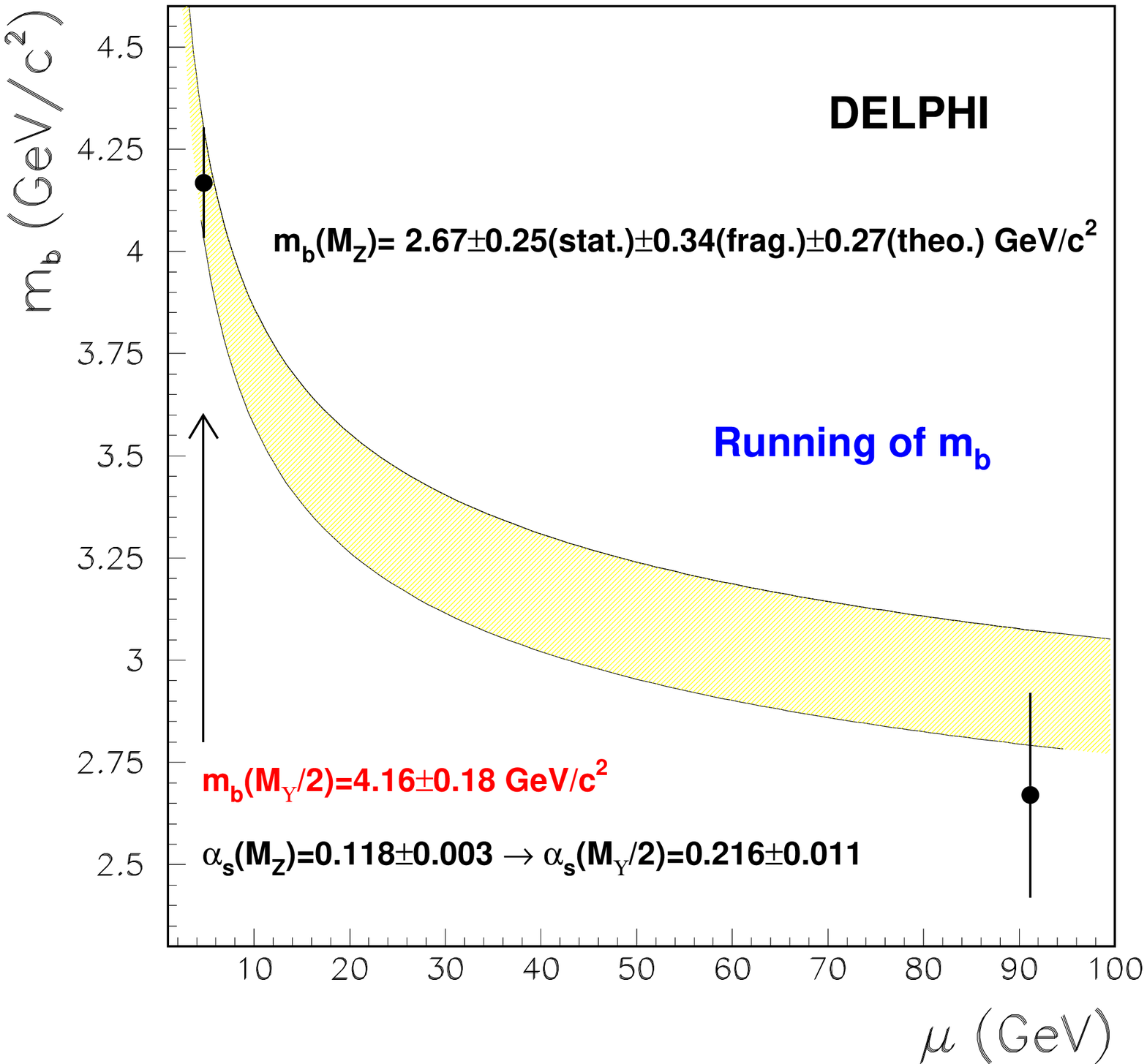}}
\end{picture}
 \begin{picture}(60,60)(0,5)
\mbox{\epsfxsize6.6cm\epsffile{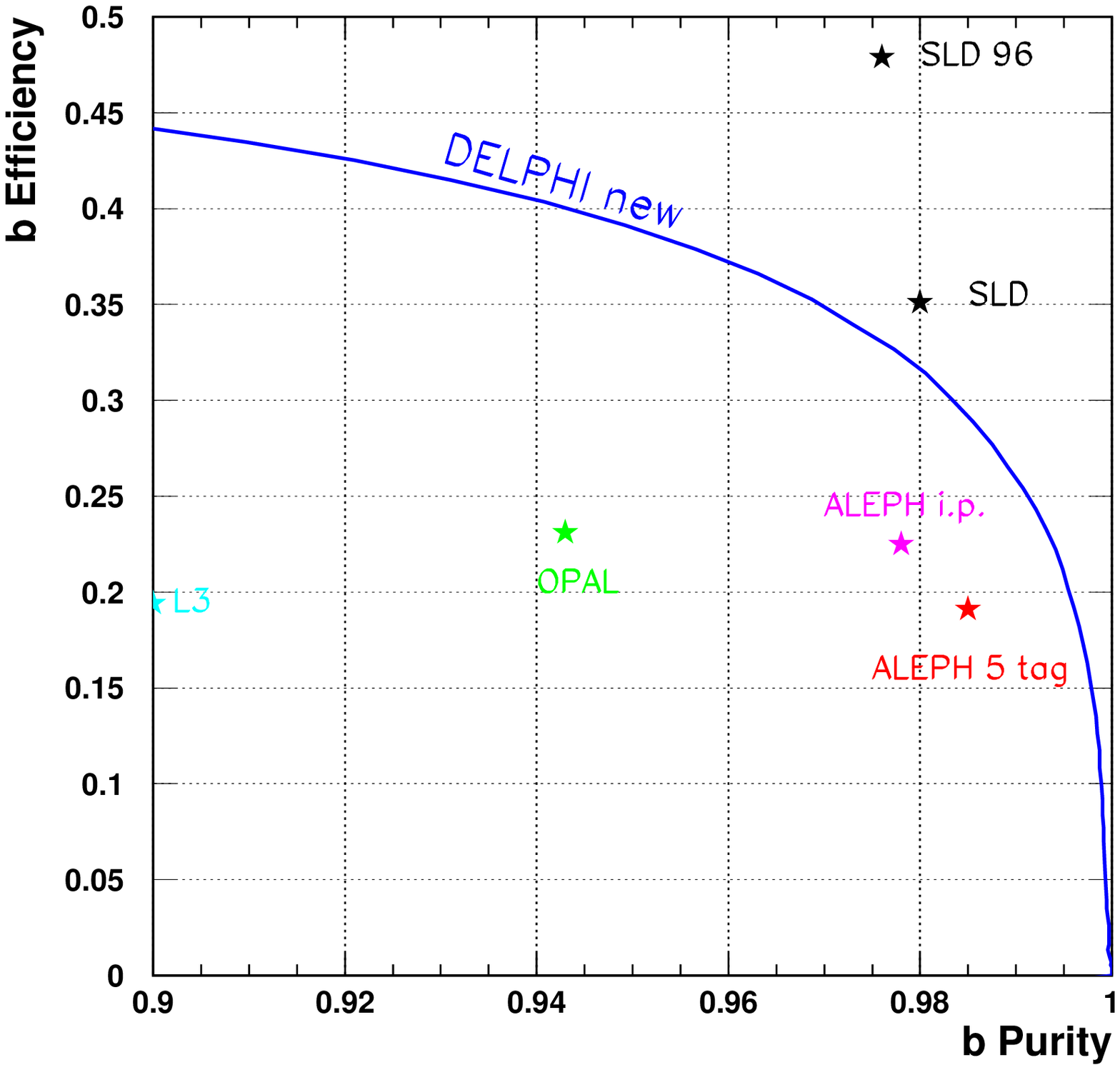}}
\end{picture}
\caption{left: b-quark mass as function of scale; right: b tagging effiency vs. purity for different experiments}
\label{effpur}
\end{figure}
heavy quarks. Comparing the measured ratio to
new next-to-leading-order (NLO) QCD calculations \cite{NLOrunningbmass}
the running $\overline{MS}$ mass has been determined to 
$m_b(M_Z)=2.67\pm0.25(stat.)\pm0.34(frag.)\pm0.27(theo.)GeV/c^2$,
to be compared to $m(b(M_\Upsilon/2)=4.16\pm0.18GeV/c^2$ from Upsilon 
spectroscopy. There is clear evidence for a running, see Fig.1a.

\section{ $b$ quark production}
\subsection{$b$  production at the Tevatron}
Both CDF and D0 find b production cross sections about 2 times as large
as NLO QCD predictions \cite{talk506}.
The shape of the transverse momentum spectrum as
well as the dependence on the c.m.s. energy is roughly described,
not however the rapidity distribution: the discrepancy gets larger in the
forward region.
\subsection{$b$ production in $Z$ decays}
The ratio $R_b$ of $Z$ decays into $b\bar{b}$-quark relative to all
hadronic (i.e. $q\bar{q}$) decays is interesting due to the large mass of the
b-quark, since an enhanced $R_b$ could indicate a Higgs-like Yukawa coupling
to masses.

There was lots of excitement because with increasing precision the world
average actually showed a discrepancy, with more than $3\sigma $ in 1995.
At last year's Warsaw Conference ALEPH 
\cite{ALEPHRb} presented an analysis with very
small errors ( $R_b=0.2161\pm0.0009\pm0.0011 $), in perfect agreement
with the SM prediction of 0.2158. 
This was less dependent on charm background and had strongly reduced
hemisphere correlations.

This year also other collaborations presented new analyses. 
Especially DELPHI [eps419] has undertaken a major
effort to reprocess all their data
with a strongly improved pattern recognition and track fitting procedure,
leading to a much cleaner reconstruction especially in dense jets.
Also the b-tagging algorithm was optimised by including the z-measurements of
the silicon vertex detector, 
vertexing, and additional variables like invariant mass
and
track rapidities. The b-tagging performances of the different detectors
are summarized in fig~1b.. Note the SLD 96 point which is due
to a new vertex detector with twice as good resolution. Unfortunately,
SLD is lacking statistics compared to the LEP Collaborations.

The  average determined by the LEP electroweak working group
\cite{EWWG}
is $0.2171\pm0.0009 $, well compatible
with the SM prediction with an accuracy of $0.4\%$. For more details see
\cite{chiara}.

\subsection{Gluon splitting}
The gluon splitting probability $f\to b\bar{b}$ is an important ingredient
in the $R_b$ measurement, constituting the largest single systematic 
uncertainty. DELPHI ( $ 0.21\pm0.11\pm0.09 $ )\cite{DELPHIgluonsplitting}
and ALEPH 
( $ 0.257\pm0.040\pm0.087)\% $ )[eps606]
have determined this parameter employing 
an analysis of b-tagged jets in four jet events, the (simple) average being
$(0.24\pm 0.09)\%$.

\subsection{$B_s$ and $B^+$ rates in b-jets at LEP}
The classical method of determining the primary $B_s$ rate $f_s$ consists of
an comparison of the integrated $B\bar{B}$ mixing $\chi$ 
in Z decays with the expectation $\chi = f_d\chi_d + f_s\chi_s$, taking
the measured $x_d$ (see below) and assuming a fast $B_s$ mixing frequency 
leading to $\chi_s=0.5$. 
The baryon contribution is estimated from lepton-$\Lambda_c$ and lepton-$\Xi$
correlations. The results of the LEP mixing working group 
\cite{LEPmixingworkinggroup} are
$f_{Baryon}=0.1062^{+0.0373}_{-0.0273}$, $f_d=f_u=0.3954^{+0.0156}_{-0.0203}$
and $f_s=0.1031^{+0.0158}_{-0.0153}$. 
DELPHI [eps451] has performed a search for a charged
fragmentation kaon accompanying
a primary $B_s$ (including the excited states $B_s^*$, $B_s^{**}$)
at high rapidity, separating
out background contributions from $B^+$ accompanied by a $K^-$.
The primary $B_s$ rate $f_S'$ has been determined to
$(12.0\pm1.4 \pm2.5 )\%$. This value \cite{weiserthesis} is smaller
than that of the contributed paper, due to a different, probably more
solid assumption about the validity of the model. With an 
estimated $B_s^{**}$ rate of $(27\pm6)\%$ this corresponds to a rate
$f_{B_s}=(8.8\pm1.0\pm1.8\pm1.0)\%$ of weakly decaying $B_s$ mesons, 
where the last error is due to the $B_s^{**}$ rate.

In the same paper [eps451] an analysis of the rate of charged versus neutral
weak B-hadrons is presented: $B(\bar{b}\to X_b^0)=(57.8\pm0.5\pm1.0)\%, 
B(\bar{b}\to X_b^+)=(42.2\pm0.5\pm1.0)\%$. Making an assumption about the
small contribution of charged $\Xi_b$ and $\Omega_b$ production
($(1.0\pm0.6)\%$) leads to $B(\bar{b}\to B^+)=(41.2\pm1.3)\%$.

ALEPH has also presented a new measurement of the b-baryon rate
of $(12.1\pm0.9\pm3.1)\%$ [eps597].

\section{Spectroscopy}
The mesons $B^0$, $B^+$ and $B_s$ are clearly established and their masses
measured. The vector meson $B^*$ is seen in its decays into $B\gamma$ and
$Be^+e^-$ [eps450]. The existence of the $L=1$ orbitally excited $B^{**}$ mesons is
established, but there is not yet a clear decomposition into the expected
2 narrow and 2 broad states. DELPHI has preliminary evidence for two 
narrow $B_s^{**}$ states decaying into $BK$ and some evidence for a radial
excitation in $B\pi^+\pi^-$, which need confirmation. The latter analysis
triggered a similar analysis in the charm sector, where a narrow
resonance in $D^*\pi^+\pi^-$ has been found [eps452].
No $L>1$ B-mesons are known.
Also searches for the beautiful charmed $B_c$-meson have been hitherto 
unsuccessful.

In the baryon sector, the $\Lambda_b$ is clearly established now by CDF
\cite{CDFLambdab},
the mass being measured to $5621\pm4\pm3$ MeV.
There are $\Sigma_b$ and $\Sigma_b^{*}$ candidates seen by DELPHI, which need
confirmation. The existence of the $\Xi_b$ is proven, but there is no 
mass measurement available.
No other b-baryon states are known: no $\Xi_b'$, $\Xi_b^*$, $\Omega_b$
or $\Omega_b^*$, also no orbital or radial excitation.

For more extensive summaries on B-spectroscopy see 
e.g.\cite{spectroscopy,had95}.

\section{Lifetimes}
Again there have been many new lifetime measurement contributions, which
are averaged by a b-lifetime working group \cite{LEPlifetimeWG} 
taking into account correlated
systematics etc. The results are:

\begin{eqnarray*}
\tau_{average} & = & 1.554\pm0.013 ps\\ 
\tau(B^0)&=&1.57\pm0.04 ps \\ 
\tau(B^+)&=&1.67\pm0.04  ps\\ \tau(B^+)/\tau(B^0) &=&1.07\pm 0.04 \\
\tau(B_s)&=&1.54\pm0.06 ps \\ \tau(b-baryon)&=&1.22\pm0.05 ps
\end{eqnarray*} 
Thus the qualitative picture remains intact: Charged B mesons live slightly
longer, the $B^0 $ and $B_s$ lifetimes are roughly the same, and the
$\Lambda_b $ has a much shorter lifetime than the mesonic states. The 
$\Lambda_b $ lifetime is correlated with a small semileptonic branching
ratio, see below. The origin of the low b-baryon lifetime is not yet
clarified.

\section{Mixing}
Second order weak interactions lead to particle-antiparticle 
oscillations between $B_0 $ and $\bar{B_0}$ and $B_s$ and $\bar{B_s}$.
They are described by a mass difference of the CP-eigenstates constructed
by the sum and difference of the original wave functions.
In the Standard Model, the mass differences are related to the Kobayashi
Maskawa matrix elements $V_{td} $ and $V_{ts} $, respectively. 
To measure the time dependence of the mixing, 
one needs to know the b-flavour at production and 
decay time to define whether a mixing occured or not, as well
as the decay length and energy to reconstruct the proper decay time.
Many different methods have been developed for this purpose. 
Fig.~\ref{bdmix} gives an overview of the available results for the $B^0 $
mass difference $\Delta m_d$, which is proportional to the oscillation
frequency: The preliminary average derived by the
LEP oscillation working group is $ 0.472\pm0.018 \,ps^{-1}$.

\begin{figure}[t]
\begin{picture}(120,60)(0,0)
\epsfxsize12.0cm\epsfysize6.0cm\epsffile{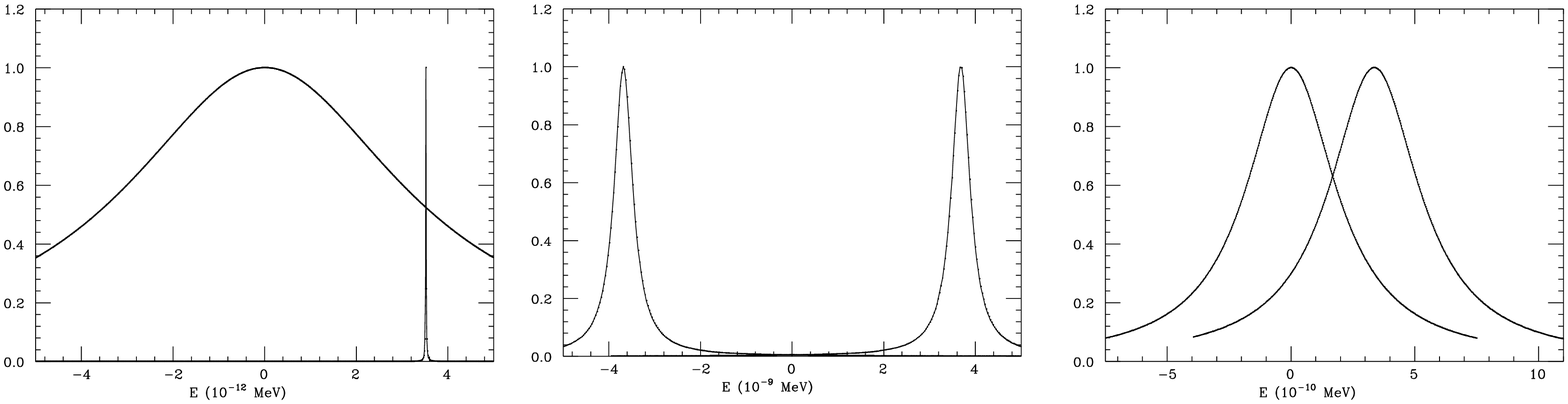}
\end{picture}
\caption{Mass excitation curves of the $K^0-\bar{K}^0$
($\Gamma _s > \Delta m \approx \Gamma_l$, $\Delta \Gamma$ large), 
$B_s-\overline{B_s}$ ($\Delta m> \Gamma$, $\Delta \Gamma$ small),  
and $ B^0-\bar{B^0}$ ($\Delta m \approx \Gamma$, 
$\Delta \Gamma$ small) systems. Curves are due to E. Golowich, Moriond 1995}
\begin{picture}(55,40)(0,0)
\mbox{\epsfxsize6.3cm\epsfysize4.3cm\epsffile{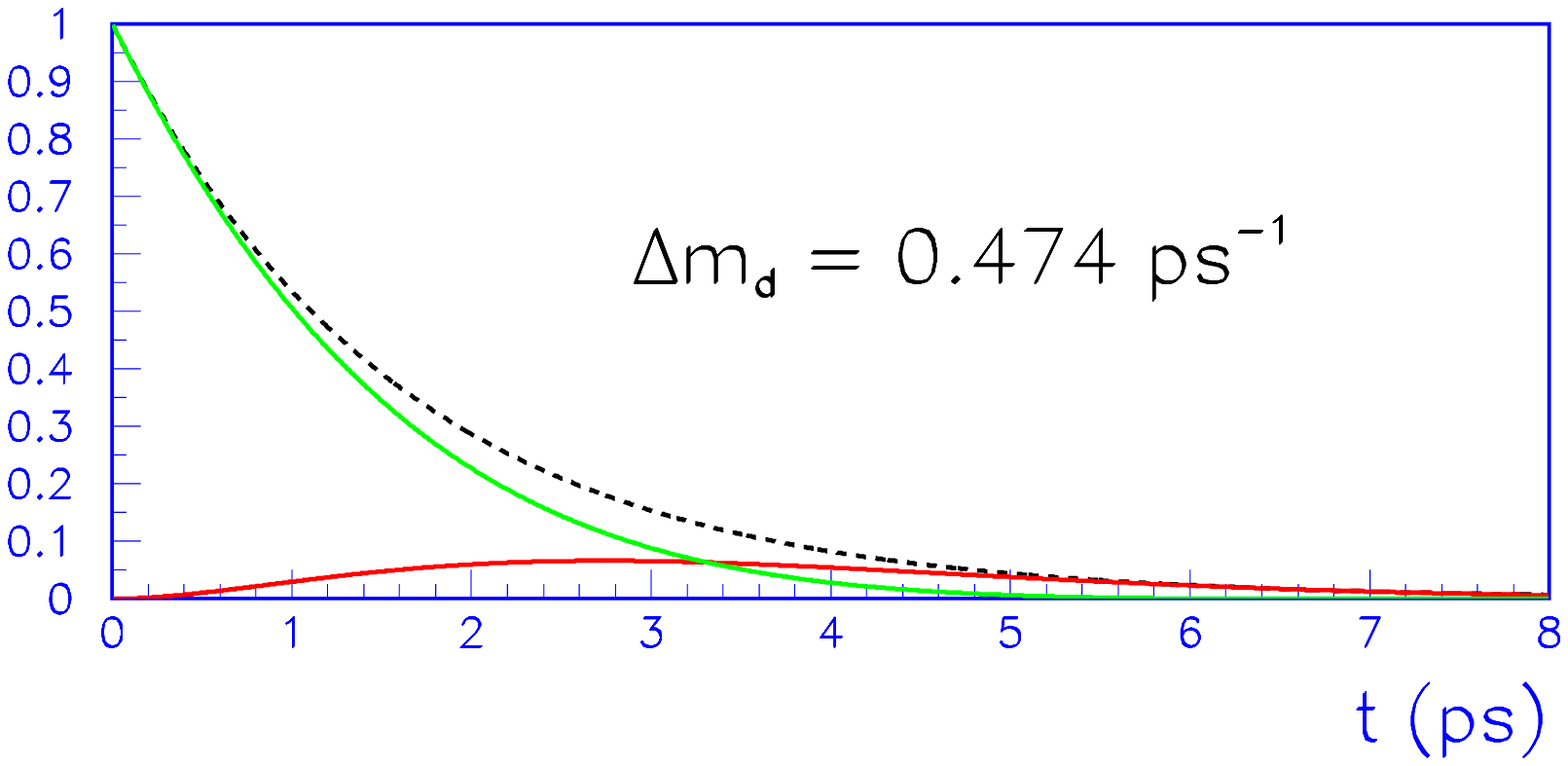}}
\end{picture}
\begin{picture}(55,40)(0,0)
\mbox{\epsfxsize6.3cm\epsfysize4.3cm\epsffile{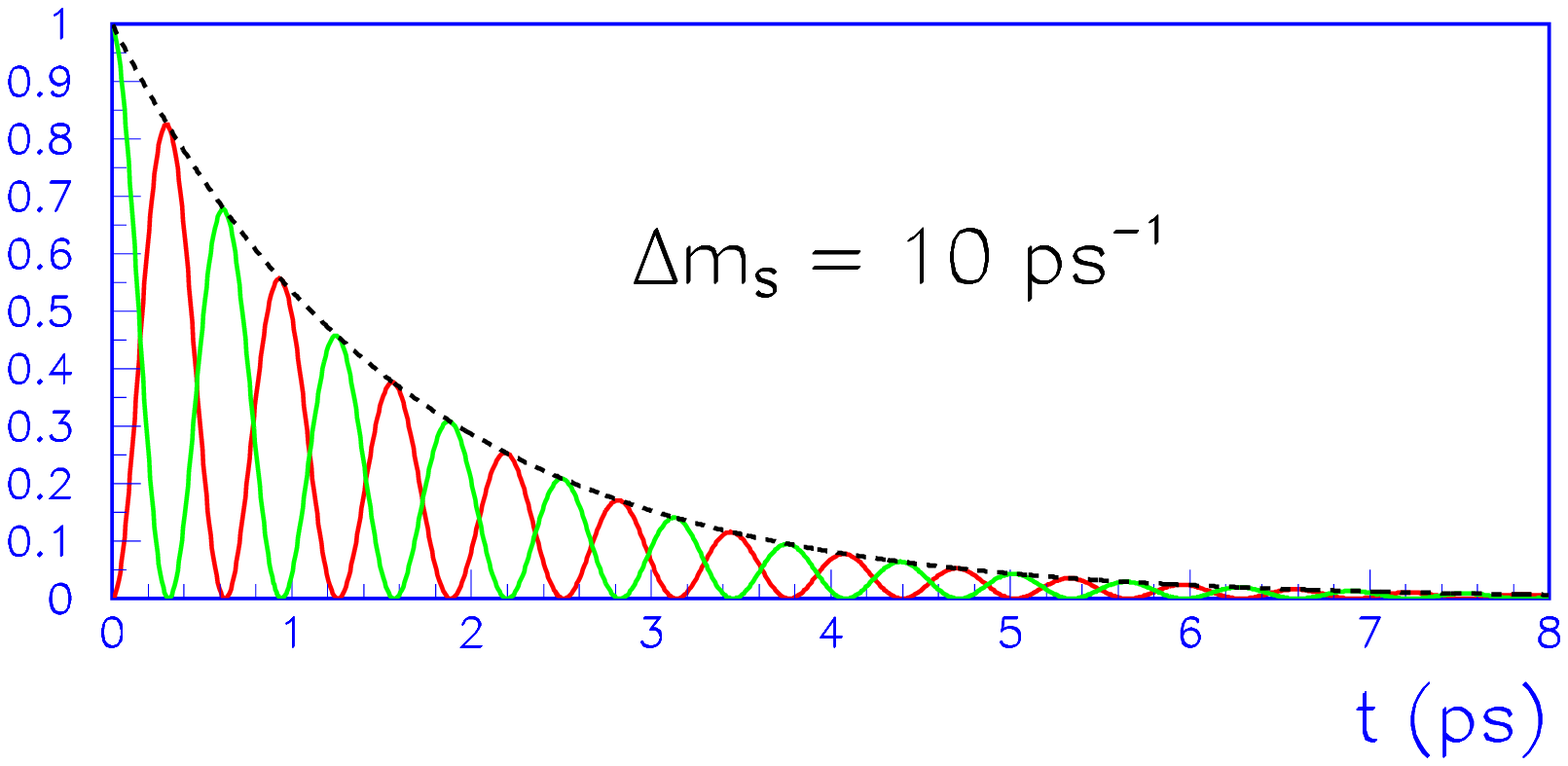}}
\end{picture}
\caption {Expected oscillation patterns for $B_d$ with 
$\Delta m = 0.474 ps^{-1}$ and $B_s$ with $\Delta m = 10 ps^{-1}$.}
\end{figure}

$B_s$ mixing proceeds much faster than $B^0$ mixing, and the time evolution
has not yet been resolved. Only a lower limit could be derived: 
\begin{figure}[p]
\centering
\begin{picture}(60,90)(13,0)
\mbox{\epsfxsize10.0cm\epsffile{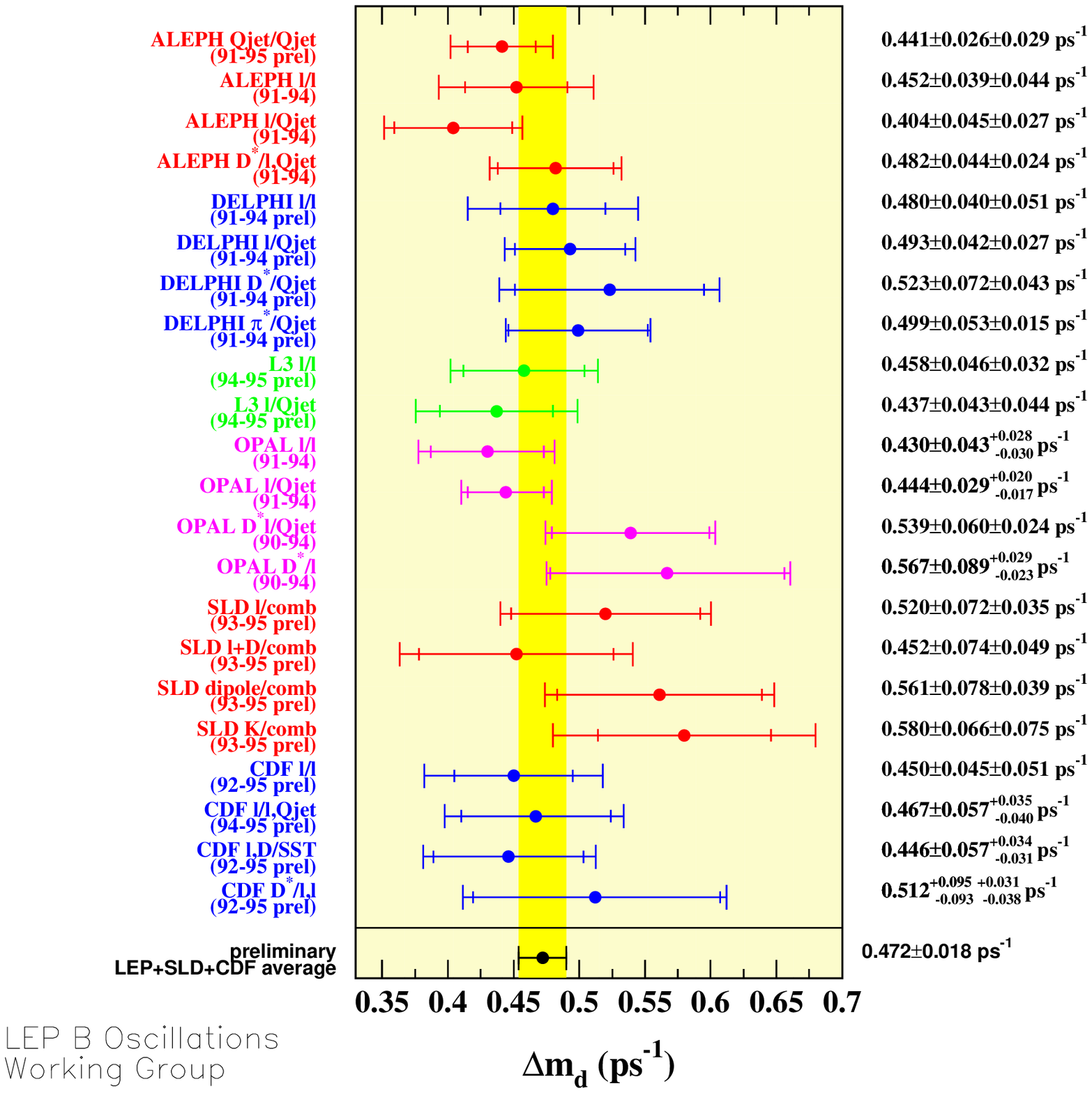}}
\end{picture}
\caption{Measurements of $\Delta m_d $} 
\label{bdmix}
\begin{picture}(60,85)(8,0)
\mbox{\epsfxsize9.0cm\epsffile{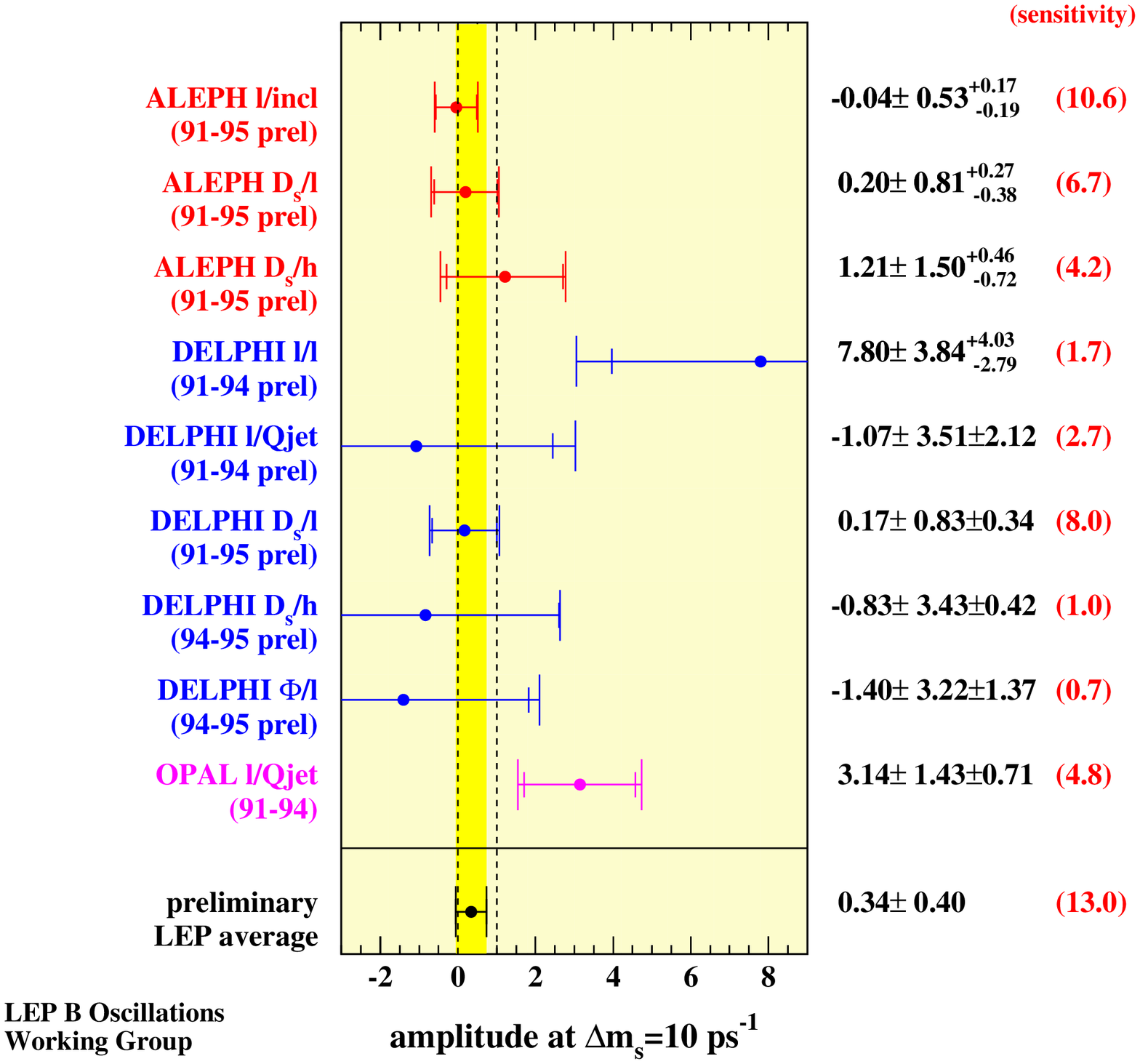}}
\end{picture}
\caption{Sensitivity on 
$\Delta m_s$}
\end{figure}
\begin{figure}[ht]
\centering
\begin{picture}(60,60)(0,0)
\epsfxsize6.0cm\epsffile{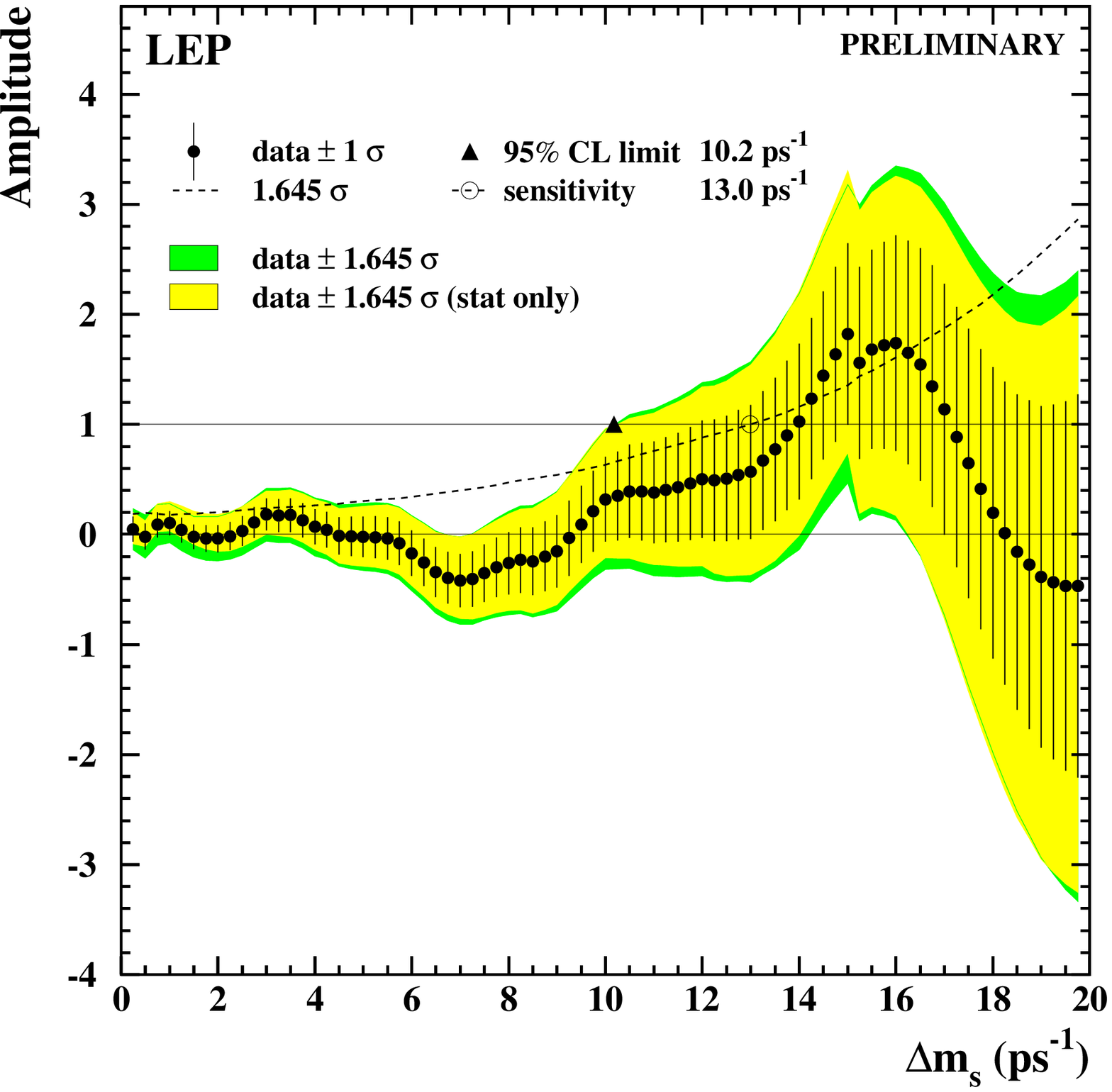}
\end{picture}
\caption{Combined $B_s$ mixing amplitude as function of $\Delta m_s$.
Physical values are 0 (i.e. no mixing with this frequency ) and 1 (data
compatible with mixing of the given frequency).}
\end{figure}
$\Delta m_s > 10.2\,ps^{-1}$ at $95 \% $ c.l. Especially noteworthy at
this conference is the new ALEPH measurement using an inclusive lepton
ansatz [eps612]. The large sensitivity came somewhat surprising, since
it was common belief that more exclusive methods with better resolution
(but less statistics) are superior. 
\section{Decays}
\subsection{Inclusive properties}
A new analysis of the mean charged multiplicity in b-hadron decays produced
at the Z by DELPHI [eps850] of 
$n(B)=4.96\pm0.03\pm0.05 $ has much smaller systematic uncertainties than
previous measurements. 

\subsection{Semileptonic decays}
The semileptonic branching ratio $B_{sl} $
is measured to be smaller than expected.
Possible explanations are large QCD corrections to $ b\to c\bar{c}s $
(which implies a large $N_c $), or a large hadronic Penguin contribution
$b\to s g$ . $N_c$ however seems to be low also (see talk of Neubert).
One should however not overlook that there are experimental mysteries:
there seem to be differences both in $ N_c$ and $ B_{sl} $ between 
the experiments using
$\Upsilon(4S) $ and $Z$ decays to produce the b-quarks, but these differences
are not as expected from the different b-hadron composition.

Both ARGUS and CLEO have performed almost model-independent analyses. They
could separate the direct $b\to l $ and the cascade $b\to c \to l$ decays
using a high energy lepton charge from the other B meson in the event,
their mean value being $10.19 \pm 0.37 \% $.
In particular, the CLEO Collaboration is sure that their result is not altered
by the discovery of ``upper vertex'' D production, as described below.
  
There is a contradiction with the LEP experiments, whose latest value
is $ B_{sl}=11.12 \pm 0.20$ 
as averaged by the electroweak working group.
It is interesting to have a closer look at this average. For the LEP HF-EW
working group $ B_{sl} $ is an auxiliary quantity in the complete electroweak
heavy flavour fit. The early measurements essentially measured 
$ R_b\times B_{sl} $, i.e. there are large correlation coefficients between
$ R_b$ and $B_{sl} $. With the current very precise values of $R_b$, the
old 1991 measurements get huge weights (and are on the high side) for the
average $ B_{sl} $ determination. This result however crucially depends on the
correlation matrix elements, including the estimates of the systematics 
correlation. Furthermore, it might be allowed to doubt that the systematics
was as well under control in 1991 as now. It should also be remarked that
the upper vertex D production is not yet included in the Monte Carlo
models used for acceptance and background calculations, and there is no
serious study yet about the possible influence. 

The value at LEP is not necessarily the same as at CLEO, due to the 
different B-hadron contribution. 
The $\Lambda_b $ has been measured to have a 
lower semileptonic branching ratio (see below), but this should lead to
a smaller value at LEP than in $ \Upsilon(4S) $ decays.

\subsection{$ \Lambda_b $ semileptonic branching ratio}
OPAL [eps153] measured the
ratio $R_{\Lambda l} = B(\Lambda_b\to \Lambda l^- X)/B(\Lambda_b \to \Lambda
X) = (7.0 \pm 1.2 \pm 0.7 ) \%$, and ALEPH [eps597] the corresponding ratio $R_{p l} = (7.8\pm1.2\pm1.4) \%$
(i.e. $\Lambda$ replaced by proton), both 
of which can be assumed to be very similar to $ B(\Lambda_b\to l X)$.
Both are significanty smaller than the average semileptonic branching
ratio. Given the apparent smaller lifetime of the $ \Lambda_b $, this
is consistent with the hypothesis of a constant semileptonic width 
for all B-hadrons.

\subsection{Wrong sign charm}
The determination of the average number of charm and anticharm
quarks per b-hadron decay is called charm counting. 
In most of the cases exactly one charm quark is produced from the b-quark
by W-emission. Only a small number of cases without c-quark is expected,
either from $b\to u $ transitions or due to loop (Penguin) processes. 
A second (anti-) charm quark is produced when the W decays into $ s\bar{c} $.
Up to recently it was thought that these two quarks always end up in a single
$\bar{D_s} $ meson. Through the measurements 
$B(B\to DX)/B(B\to\bar{D}X) =0.100\pm0.026\pm0.016$
(CLEO) [eps383],
$ B(B^{0,-}\to D^0\bar{D^0}X, D^0D^-X, \bar{D^0}D^+X) =12.8\pm2.7\pm2.6$ 
(ALEPH) \cite{ALEPHwrongsigncharm} and
$ B(B^{0,-}\to D^{*+}D^{*-}X) =1.0\pm0.2\pm0.3$ (DELPHI)
\cite{DELPHIwrongsigncharm}
we know that this is not true. CLEO has performed this measurement by
analysing angular correlations of D-mesons with high momentum leptons.
CLEO also has observed four exclusive double charm decay 
modes and has placed limits on three others
[eps337].
The observed large rates including $D^* $ mesons suggest that the wrong sign D 
mesons have a very soft spectrum in the B cms. CLEO also has searched for
resonances (especially the $J^P=1^+ D_{s1}^+(2536) $) in the upper vertex
$D^*K $ spectra, but didn't find any enhancement[eps384]. 
From their upper limit 
$ B(B\to D_{s1}^+X)<0.95\%$ at $95\%c.l.$ one can deduce that the 
axial vector coupling constant $ f_{D_{s1}^+} $ is at least a factor 2.5
lower than that of the pseudocalar $ D_s^+$.

\subsection{Charm Counting}
Classical charm counting experiments consist in 
measuring the rates of the weakly decaying D-hadrons in selected b-events.
The published values differ quite a bit:
CLEO: $ N_c = 111.9 \pm 1.8 \pm 2.3 \pm 3.3\% $ \cite{CLEOnc}, 
ALEPH: $ N_c = 123.0 \pm 3.6 \pm 3.8\pm5.3 \% $ \cite{ALEPHnc},
OPAL: $ N_c=106.1\pm4.5\pm6.0\pm3.7\%$ \cite{OPALnc}, where the last error is
due to D branching ratios, largely correlated between the experiments.
OPAL measures comparatively small $D^0$ and $ D^+$ rates.
A main difference between the experiments however are assumptions made about
the unmeasured $\Xi_c $ contribution, which is set to 0 in the case of OPAL,
whereas ALEPH estimates it to be $ 6.3\pm2.1\%$. 
Accepting this last estimate and including also DELPHI's measurement of
$D^0 $ and $D^+ $ rates \cite{DELPHID} the averaged result is
$ N_c = (120.2 \pm 4.0 (stat+syst) \pm 5.3 (BR)) \% $.

Two alternative methods to determine the fraction of b-decays into 0,1 and
2 charmed hadrons have been suggested by the DELPHI Collaboration
\cite{kluit,DELPHIdoublecharm}:
An analysis of the hemisphere b-tagging probability distribution in terms
of Monte Carlo expectations of the three components delivers the result
$ B(b\to0c)=4.4\pm 2.5 \%$, $B(b\to2c)=16.3\pm4.6\%$, and 
$ N_c=116.3\pm4.5\%$. In another ansatz correlations of identified charged
kaons with inclusively reconstructed D mesons are analysed. A fit of the
transverse momentum spectra of same sign and opposite sign K pairs results
in $B(b\to2c)=17.0\pm3.5\pm3.2\% $ and 
$ B(b\to \bar{D}D_sX)/B(b\to2c)=0.84\pm0.16\pm0.09 $. 
Large rates of $ b\to s g$, as proposed by Kagan\cite{kagan}, would show
up as extra source of charged kaons, especially visible at high momentum in 
the B c.m.s.. This is not seen in DELPHI, and an upper limit 
$B(b\to s g) <5\%$ at $ 95\%$ c.l. is derived. The SLD Collaboration
\cite{SLDkaon}
however finds a small excess in the kaon $ p_T$ spectrum, when they
demand that the tracks form a good single vertex (to enhance b-decays
without secondary charm decay), but they did not present a numerical 
analysis yet.

In their wrong sign charm paper CLEO [eps383] 
also derives the numbers
$ B(b\to s g) = 0.2\pm 4.0\% (<6.8\%$ at $90\%$c.l.),
$ B(b\to c\bar{c}s) = 21.9\pm 3.6\%$ and $n_c=120.4\pm3.7\%$.

Although there still are some discrepancies at the 2 sigma level, and there
were controversial discussions on many of the analyses involved, it seems
that there is not a serious $N_c$ problem any more.
Combining all the numbers leads to $ N_c=117.6\pm2.3\%$.
 
\subsection{$ B^+$ branching fractions}
Although many inclusive branching ratios have been measured at ARGUS and
CLEO, most of them are B-inclusive and do not distinguish between 
$ B^+$, $ B^0$, $\bar{B^0}$, and $B^-$. DELPHI [eps473] has presented a 
feasibility study of a method
to enrich inclusively $ B^+$ mesons and to measure $ \pi^+$, $ \pi^-$,
$ K^+$, $ K^-$, $ e^+$, $ e^-$, and $ \mu^+$ and $ \mu^-$ rates as function
of the momentum in the B-meson c.m.s.. Also the rates of different
D-hadron species in $ B^+$ decays are largely unknown and can be addressed
with this method. Furthermore the method can be modified to a 
flavour-specific ($ b$-$ \bar{b}$) study.
\begin{figure}[ht]
\begin{picture}(60,60)(0,15)
\epsfxsize12.0cm\epsffile{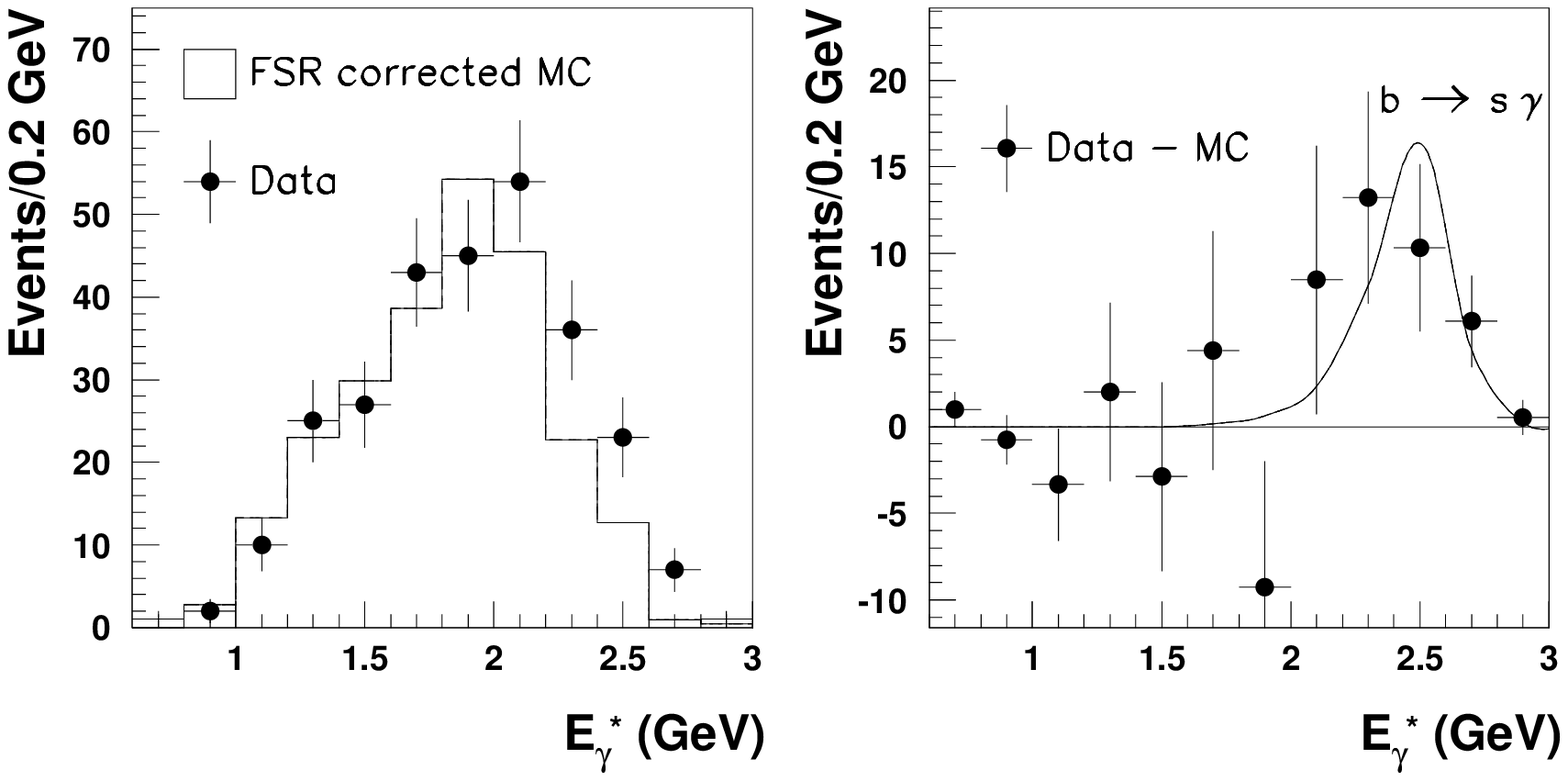}
\end{picture}
\caption{ALEPH signal of $ B\to s \gamma $ }
\end{figure}

\subsection{$ V_{cb}$}
There is not much news on $V_{cb}$. Mean values from different
reactions are \cite{drell} 
$B\to D^{*}l\nu$: $ (38.7\pm3.1)\cdot10^{-3}$,
$B\to Dl\nu$: $ (39.4\pm5)\cdot10^{-3}$,
$\Upsilon(4S)$ inclusive: $ (38.7\pm2.1)\cdot10^{-3}$,
$Z^{0}$ inclusive: $ (40.6\pm2.1)\cdot10^{-3}$. 
The last two correlated values can be combined into 
$ (39.9\pm2.2)\cdot10^{-3} $, leading to an overall average value of
$(39.5\pm 1.7)\cdot 10^{-3}$.
A limiting factor in exclusive and semiexclusive analyses is the bad
knowledge of $ D^{**}$ and nonresonant $D\pi$ production. 

\subsection{ $ V_{ub}$ }
Three measurements of $V_{ub}$ are available: CLEO's lepton
endpoint spectrum $(3.1\pm0.8)\cdot10^{-3}$, CLEO's exclusive
$B\to\pi/\rho l \nu$ value $3.3\pm0.3^{+0.3}_{-0.4}\pm 0.7)\cdot 10^{-3}$ 
and ALEPH's neural network analysis \cite{ALEPHVub} with 
$ ( 4.3\pm0.6\pm0.6  )\cdot 10^{-3}   $.
All of them are strongly model dependent, however in different ways.
The good agreement between the numbers is thus comforting.
 
\subsection{$ b\to s \gamma $}

The electomagnetic penguin $ b\to s \gamma$ has now also been observed by the
ALEPH Collaboration at a rate of
$ B(b\to s \gamma) = (3.29\pm0.71\pm 0.68) \cdot 10^{-4}$. 
Averaged with the 1994 CLEO result this corresponds to a new mean value
of $ B(b \to s \gamma) = (2.578\pm0.57)\cdot 10^{-4}$. 
New next to leading order calculations \cite{theorybtosgamma} are 
$ (3.28\pm 0.31)\cdot 10^{-4}$ and 
$ (3.48\pm 0.33)\cdot 10^{-4}$, slightly larger than the measured value.
One has to wait for an updated CLEO number with smaller errors. 

\subsection{Hadronic penguins}
CLEO, ALEPH and DELPHI have observed a number of still very small signals on  
exclusive charmless final states with branching ratios in the order of
$ 10^{-5}$. A new CLEO analysis [eps334] show that 
the penguin contributions (e.g. $ B\to K \pi $) might be larger than expected
compared to $ b\to u $ transitions (like $ B\to \pi\pi$). Especially this
latter result is worrisome for the prospects of measuring the CKM-phase
$ \gamma$ at future b-factories from the decay $B\to \pi^+\pi^-$. Particle
identification becomes more and more important! CLEO also has evidence for
exclusive final states including $ \omega$ and $ \phi$ mesons 
[eps335].

\subsection{B decays involving $ \eta'$ } 
CLEO finds relatively large rates of charmless decays involving
$ \eta'$ mesons [eps333]: 
$ B(B^\pm\to\eta'K^\pm)=(7.1^{+2.5}_{-2.1}\pm0.9)\cdot 10^{-5}$,
$ B(B^0\to\eta'K^0)=(5.3^{+2.8}_{-2.2}\pm1.2)\cdot 10^{-5}$ [eps333], and
inclusively $ B(B\to\eta'X)=(6.2^\pm1.6\pm1.1)\cdot 10^{-4}$ (with
$ 2.0<p(\eta')<2.7 GeV$) [eps332].
One might speculate whether there is a larger than expected $ c\bar{c}$
or glueball component in the $ \eta'$ wave function. 
\begin{figure}[t]
\centering
\begin{picture}(60,60)(0,0)
\epsfxsize6.0cm\epsffile{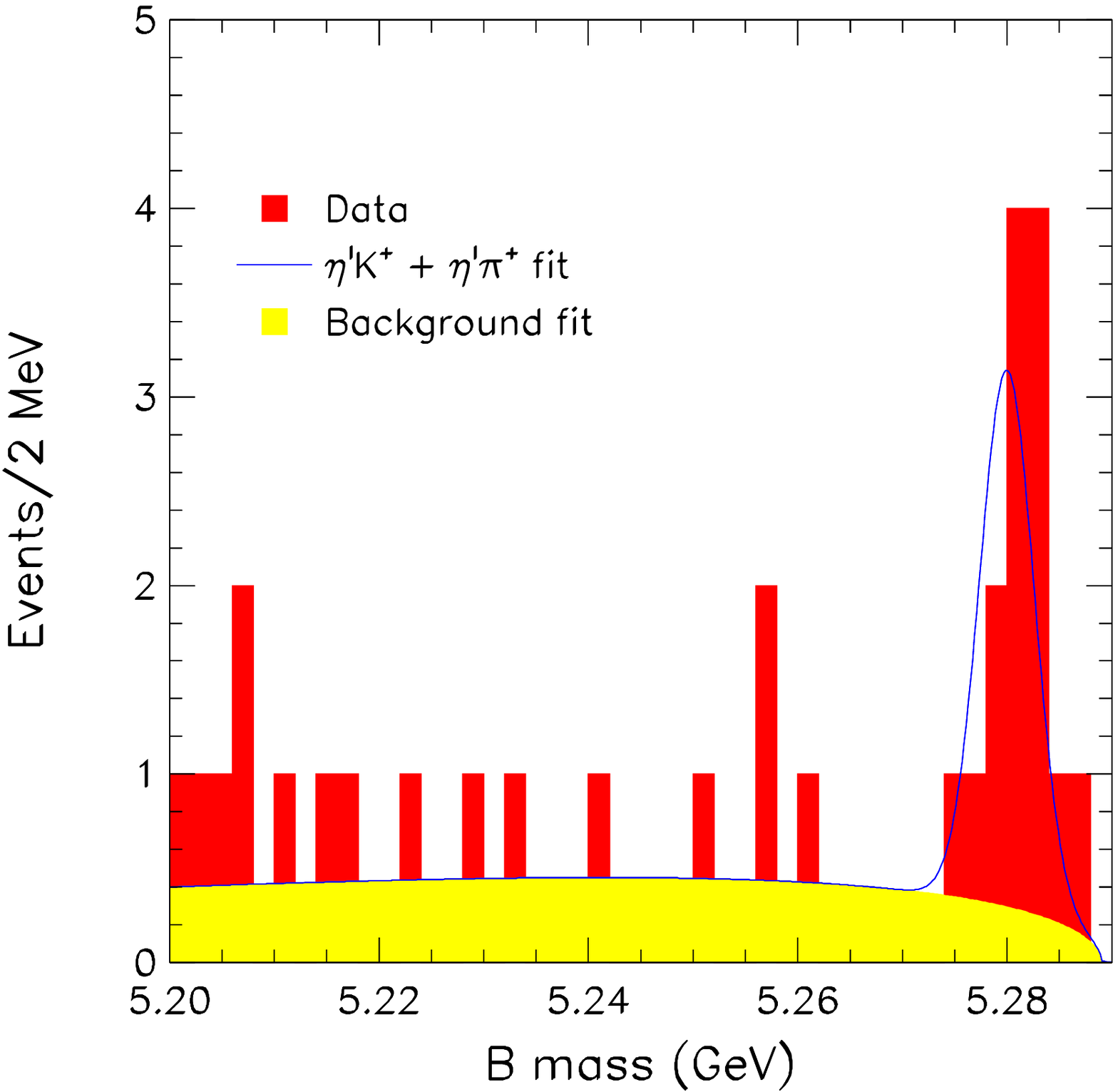}
\end{picture}
\caption{CLEO signal of $ B\to \eta' K $ }
\end{figure}
In this respect also another CLEO analysis is of interest: The measurement
of the electromagnetic form factors of the $\pi^0$,$ \eta$ and $ \eta'$,
as pioneered by 
TPC/2$ \gamma$ \cite{TPC} and 
CELLO \cite{CELLO} , now
is measured up to such high $ Q^2$ that a very remarkable 
qualitative statement can be made [eps703]:
The $ Q^2$ dependence of the $\eta'$ form factor cannot
be described simultaneously at low and high $ Q^2$ 
with the same formalism as the $ \pi^0$ and $ \eta$ mesons. This might
be another clue that there is something more than just light quarks 
inside the $ \eta'$.

CLEO finds much smaller branching ratios for decays incolving $ \eta$
mesons, as predicted by Lipkin as interference effect between creating
the $ \eta$ and $\eta'$ by their $u\bar{u},d\bar{d}$ component and
their $ s\bar{s}$ component.

\section{Other puzzles and open questions, further studies}
\subsection{Interference}
CLEO [eps339] has measured the
interference sign between colour suppressed and 
colour allowed amplitudes (which are closely connected to internal and
external spectator diagrams) to be positive and consistent with equal 
in the six
final states $D\pi,D\rho,Da_1$ and $D^*\pi$,$ D^*\rho$ and $D^*a_1$. From this
they would expect an up to 
$ 15\%$ 
larger $ B^0$- than $ B^+$-lifetime, in
contradiction to experiment. This must mean that this interference pattern
is not typical for the whole set of hadronic B-decays. 
\subsection{$ B^+,B^0$ production rates in $\Upsilon(4S)$ decays }
It is worrying to observe that the $ B^+$ to $B^0$ production ratio in
$ \Upsilon(4S)$ decays still is known only very badly:   
$ f_{+-}/f_{00}=1.21\pm0.12\pm0.17$ (CLEO alone), or
$ f_{+-}/f_{00}=1.074\pm0.129$ (CLEO, with world average lifetimes).
Most analyses assume that the ratio is 1, but due to phase space
effects a non-zero value is not excluded. If the charged and neutral
lifetime are really different, and the production ratio is not 1, there could
be quite some surprises in CLEO-LEP comparisons. However, a larger 
semileptonic branching ratio at LEP than at the $ \Upsilon(4S)$ is not
achievable with current input data.  
\subsection{Search for CP-violation}
Both OPAL[eps162]
and DELPHI[eps449]
have searched for CP-violating effects and established limits on
$Re(\epsilon_{b})$.
\subsection{CKM matrix fits}
A combination of B-mixing results, $V_{ub}/V_{cb}$ and
CP-violation parameters from $K^0$ decay shows clear evidence of a non-trivial
unitarity triangle. With some input from theory two angles of the 
triangle can already be determined with quite good precision \cite{roudeau}:
$sin 2\alpha = -0.10 \pm 0.40 $ and
$sin 2 \beta = 0.68 \pm 0.10$.   

\section{Summary and Outlook}
There is a bright future for b-physics in front of us - it will dominate
the experimental high energy physics scene between 2000 and the LHC
startup in 2007 or so.
At LEP the final analyses with optimised algorithms are being prepared.
SLC will hopefully get higher statistics before it will be shut down. 
CLEO has upgraded the detector, and CESR's luminosity will continue 
to improve. In 1999 the b-factory detectors BABAR and BELLE will start to take 
data. CDF and D0 get upgraded and will take data at much higher luminosity
at the Tevatron, HERA-B at DESY will enter the scene, and finally LHCB
and perhaps BTEV will be able to do many analyses with huge precision.

In a few years from now many rare tree and penguin decays will be known, the
$ B_c$ will be discovered, time-dependence of $B_s$ mixing detected, and
CP-violation observed in many channels. Then we will probably laugh about the
few 2 sigma discrepancies that we have to deal with now.

\section{Acknowledgements}
Many thanks to the ALEPH, CDF, CLEO, DELPHI, D0, L3, OPAL and SLD 
Collaborations, the LEP Electroweak Heavy Flavour, B-lifetime and
B-oscillation working groups, and in particular
P. Antilogus, P. Drell, L. Gibbons, T. Hessing, C. Kreuter, J. Kroll, 
C. Mariotti, P. Maettig, K. Moenig, M. Neubert, P. Roudeau, E. Thorndike, 
I. Tomalin and S. Willocq for their help in preparing the review.

%


\begin{thebibliography}{999}
\bibitem{roudeau} P. Roudeau, LAL97-96, to appear in Nuovo Cimento
\bibitem{schneider} O. Schneider, CERN-PPE/97-143
\bibitem{drell} P. Drell, CLNS 97/1521
\bibitem{neubert} M. Neubert, talk 19, these proceedings and CERN-TH-98-2
\bibitem{feindt} M. Feindt,http://wwwinfo.cern.ch/$\tilde{\mbox{ }}$feindt 
\bibitem{DELPHIrunningbmass} DELPHI Coll., CERN-PPE-97-141
\bibitem{talk514} J. Fuster, talk 514, these proceedings  
\bibitem{NLOrunningbmass} G. Rodrigo, Nucl. Phys. Proc. Suppl. 54A (1997) 60;
\newline G. Rodrigo, M. Bilenky, A. Santamaria, Phys. Rev. Lett. 79 (1997) 193;
\newline W. Bernreuther, A. Brandenburg, P. Uwer, Phys. Rev. Lett. 79 (1997) 189;
\newline P. Nason, O. Olcari, Phys. lett. B407 (1997) 57 
\bibitem{talk506} R. Jesik, talk 506, these proceedings  
\bibitem{ALEPHRb} ALEPH Coll., R. Barate et al., Phys.Lett.B401 (1997) 150, 
Phys.Lett.B401 (1997) 163
\bibitem{EWWG}The LEP Experiments and the LEP Electroweak Working Group, CERN-PPE-97-154
\bibitem{chiara} C. Mariotti, talk 707, these proceedings
\bibitem{DELPHIgluonsplitting}  DELPHI Coll., P. Abreu et al., Phys.Lett. B405 (1997) 202 
\bibitem{LEPmixingworkinggroup} LEP Oscillation Working Group,
http://www.cern.ch/LEPBOSC/
\bibitem{weiserthesis}
C. Weiser, Ph.D. thesis, University of Karlsruhe, in print
\bibitem{CDFLambdab} CDF Coll., F. Abe et al.,Phys. Rev. D 55 (1997) 1142
\bibitem{spectroscopy} V. Canale, talk 103, these proceedings
\bibitem{had95} M. Feindt, Procs. Hadron 95, Manchester, UK, World Scientific,
  p.241  \newline M. Feindt, Acta Phys. Pol. B 28 (1997) 789
\bibitem{LEPlifetimeWG} 
LEP Lifetime Working Group, 
http://wwwinfo.cern.ch/$\tilde{\mbox{ }}$claires/ \linebreak
lepblife.html 
\bibitem{ALEPHwrongsigncharm} ALEPH Coll., contributed paper to ICHEP'96 pa05-060, Warsaw 1996 
\bibitem{DELPHIwrongsigncharm} DELPHI Coll., contributed paper to ICHEP'96 pa01-108, Warsaw 1996 
\bibitem{CLEOnc} CLEO Coll., L. Gibbons et al., CLNS 96/1454, hep-ex/97030006
\bibitem{ALEPHnc} ALEPH Coll., D. Buskulic et al., Z. Phys. C 69 (1996) 585
\bibitem{OPALnc} OPAL Coll., G. Alexander et al., Z. Phys. C 72 (1996) 1
\bibitem{kluit} P. Kluit, talk 906 at this conference
\bibitem{DELPHIdoublecharm} DELPHI Coll., CERN-EP-98-007
\bibitem{DELPHID} DELPHI Coll., contributed paper to ICHEP'96 pa01-058  
\bibitem{kagan} A. Kagan, talk 512 at this conference
\bibitem{SLDkaon} D. Jackson, talk 511 at this conference
\bibitem{ALEPHVub} ALEPH Coll., contributed paper to ICHEP'96 pa05-59, Warsaw 1996  
\bibitem{theorybtosgamma} K.G. Chertyrkin, M. Misiak and M. Munz, Phys. Lett. B400 (1997) 206;
\newline
A.J.Buras, A. Kwiatkowski and N. Pott, TUM-HEP-287/97
\bibitem{talk510} D. Miller, talk 510 at this conference
\bibitem{TPC} TPC/2$\gamma$ Coll., H. Aihara et al., Phys. Rev. Lett. 64 (1990) 172
\bibitem{CELLO} CELLO Coll., H.-J. Behrend et al., Z. Phys. C49 (1991) 401 
\end{thebibliography}
\end{document}